\documentclass{ws-ijmpa-blank-new}
\usepackage[super,compress]{cite}
\usepackage{graphicx}

\usepackage{amssymb,epsf,latexsym,amssymb,amsmath,mathrsfs}
\usepackage[english]{babel}

\newcommand{\beq}{\begin{equation}}
\newcommand{\eeq}{\end{equation}}
\newcommand{\beqa}{\begin{eqnarray}}
\newcommand{\eeqa}{\end{eqnarray}}
\newcommand{\bsubeqs}{\begin{subequations}}
\newcommand{\esubeqs}{\end{subequations}}
\newcommand{\half}{{\textstyle \frac{1}{2}}}

\makeatletter
\@addtoreset{equation}{section}
\renewcommand{\theequation}{\thesection.\arabic{equation}}
\makeatother

\begin{document}
\markboth{F.R. Klinkhamer, G.E. Volovik}
{Dynamic cancellation of a cosmological constant...}

%
\catchline{}{}{}{}{}
%

\title{\vspace*{-9mm}
Dynamic cancellation of a cosmological constant and
approach to the Minkowski vacuum}

\author{F.R. Klinkhamer}

\address{Institute for Theoretical Physics, Karlsruhe Institute of
Technology (KIT),\\ 76128 Karlsruhe, Germany\\
frans.klinkhamer@kit.edu}

\author{G.E. Volovik}

\address{Low Temperature Laboratory, Department of Applied Physics,
Aalto University,\\ 
PO Box 15100, FI-00076 Aalto, Finland,\\
and\\
Landau Institute for Theoretical Physics RAS,\\ Kosygina 2,
119334 Moscow, Russia\\
volovik@ltl.tkk.fi}

\maketitle


\begin{abstract}
\\
The $q$-theory approach to the cosmological constant problem
is reconsidered. The new observation is that
the effective classical $q$-theory gets modified
due to the backreaction of quantum-mechanical particle production
by spacetime curvature.
Furthermore,  a Planck-scale cosmological constant is added to 
the potential term of the action density, in order to 
represent the effects from zero-point energies and phase transitions.
The resulting dynamical equations of 
a spatially-flat Friedmann--Robertson--Walker universe
are then found to give a steady approach to the Minkowski vacuum,
with attractor behavior for a finite domain of
initial boundary conditions on the fields.
The approach to the Minkowski vacuum is slow and gives rise
to an inflation-type increase of the particle horizon.
\\
\end{abstract}
\hspace*{0mm}
{\footnotesize
\emph{Journal}: \emph{Mod. Phys. Lett. A} \textbf{31}, 1650160 (2016)
\vspace*{.25\baselineskip}\newline
\hspace*{5mm}
\emph{Preprint}: arXiv:1601.00601 
}
\vspace*{-5mm}\newline
\ccode{PACS Nos.: 03.70.+k, 98.80.Cq, 98.80.Es}



\section{Introduction}
\label{sec:Intro}

Several years ago, we proposed a particular approach to the
cosmological constant problem~\cite{Weinberg1988}, whose
motivation relies on thermodynamics and Lorentz invariance
and which goes under the name of $q$-theory~\cite{KlinkhamerVolovik2008a}.
The basic idea of $q$--theory is to give
the proper \emph{macroscopic} description of the
Lorentz-invariant quantum vacuum where the gravitational
effects of a (Planck-scale) cosmological constant $\Lambda$
have been cancelled dynamically
by appropriate \emph{microscopic} degrees of freedom.
In general,  there are one or more of these
vacuum variables (denoted by $q$, with or without
additional suffixes) to characterize the thermodynamics
of this \emph{static} physical system in equilibrium.
Several realizations of  $q$--theory have been given, but
the most elegant is the one with $q$ arising from
a four-form field strength $F$ (details and references
are given below).
The outstanding issue is the \emph{dynamics},
namely, how the cosmological constant $\Lambda$ is
cancelled dynamically and the equilibrium state approached.

In a follow-up paper~\cite{KlinkhamerVolovik2008b},
we established the dynamic relaxation of
the vacuum energy density to zero, \emph{provided} the chemical
potential $\mu$ has already the equilibrium value $\mu_0$
corresponding to the Minkowski vacuum.
But, then, the cosmological constant problem is replaced
by another problem~\cite{KlinkhamerVolovik2010,KlinkhamerVolovik2016b}, namely, why does the constant $\mu$ have the ``right'' value $\mu_0$.

Here, we discuss how quantum effects can modify the
classical $q$-theory and give rise to the decay of the
vacuum energy density (i.e., decay of the
effective chemical potential). Related work on
vacuum energy decay has been presented in
Refs.~\citen{Kofman1997,Polyakov2008,Polyakov2010,KrotovPolyakov2011,%
Polyakov2012,Pimentel2015,Klinkhamer2012,Felice2012},
but the feedback on $q$-theory has not been considered in detail.

The present paper is self-contained but somewhat short on
the underlying ideas of $q$-theory.
More details on the condensed-matter-physics motivation of $q$-theory
can be found in a companion paper~\cite{KSV2016}, which
contains a general discussion of the issue of vacuum-energy relaxation.

Throughout, we use natural units with $c=1$ and $\hbar=1$,
unless stated otherwise. We also take the metric signature $(-+++)$.

\section{Four-form-field-strength realization of classical $\boldsymbol{q}$-theory}
\label{sec:Realization}

Start by neglecting the energy exchange between vacuum and matter.
Then, the dynamics is described by the following classical
action~\cite{KlinkhamerVolovik2008b}:
\bsubeqs\label{eq:EinsteinF-all}
\beqa
&&
I=
\int_{\mathbb{R}^4}
\,d^4x\, \sqrt{-g}\,\left(\frac{R}{16\pi\, G(q)} +\epsilon(q)
+\mathcal{L}^\text{\,SM}(\psi)\right) \,,
\label{eq:actionF}\\[2mm]
&&
q^2 \equiv- \frac{1}{24}\,
F_{\kappa\lambda\mu\nu}\,F^{\kappa\lambda\mu\nu}\,,\quad
F_{\kappa\lambda\mu\nu}\equiv
\nabla_{[\kappa}A_{\lambda\mu\nu]}\,,
\label{eq:Fdefinition}\\[2mm]
&&
F_{\kappa\lambda\mu\nu}=q\sqrt{-g} \,\epsilon_{\kappa\lambda\mu\nu}\,,\quad
F^{\kappa\lambda\mu\nu}=q \,\epsilon^{\kappa\lambda\mu\nu}/\sqrt{-g}\,,
\label{eq:Fdefinition2}
\eeqa
\esubeqs
where Eqs.~\eqref{eq:Fdefinition} and
\eqref{eq:Fdefinition2} give a particular realization of the vacuum
$q$-field in terms of the 4-form field strength $F$ from a three-form
gauge field $A$~\cite{DuffNieuwenhuizen1980,Aurilia-etal1980,Hawking1984,%
Duff1989,DuncanJensen1989,BoussoPolchinski2000,Aurilia-etal2004,Wu2008}.
The symbol $\nabla_\mu$ in \eqref{eq:Fdefinition} denotes the
covariant derivative and a pair of square brackets
around spacetime indices stands for complete anti-symmetrization.
The symbol $\epsilon_{\kappa\lambda\mu\nu}$ in \eqref{eq:Fdefinition2}
corresponds to the Levi--Civita symbol, which makes $q$  a pseudoscalar.
Note that $q$ in the action \eqref{eq:actionF} is a pseudoscalar
but not a fundamental pseudoscalar, $q=q(A,\,g)$. The fundamental
fields of the theory considered are the gauge field $A_{\mu\nu\rho}(x)$,
the metric $g_{\mu\nu}(x)$, and the generic matter field $\psi(x)$.

In the action \eqref{eq:actionF},
$\mathcal{L}^\text{\,SM}(\psi)$  is the Lagrange density of the
standard-model matter fields $\psi$.
The parameters of the matter action may depend, in principle,
on the vacuum variable $q$, $\mathcal{L}^\text{\,SM}$ $=$ $\mathcal{L}^\text{\,SM}(q,\psi)$.
But, here, we neglect this $q$ dependence
of the standard-model parameters and allow  only
for a $q$ dependence of the gravitational coupling, $G=G(q)$.

The variation of the action \eqref{eq:actionF} over the three-form
gauge field $A$ gives the generalized Maxwell equation for the
four-form field strength $F$,
\begin{equation}
\nabla_\nu \left(\sqrt{-g} \;\frac{F^{\kappa\lambda\mu\nu}}{q}
\left[
\frac{d\epsilon(q)}{d q}+\frac{R}{16\pi} \frac{dG^{-1}(q)}{d q}
\right]   \right)=0\,.
\label{eq:Maxwell}
\end{equation}
In the spatially-flat ($k=0$) Friedmann--Robertson--Walker (FRW) universe
with comoving coordinates, this can be written as
\begin{equation}
\partial_{t} \left(
\frac{d\epsilon(q)}{d q}
-\frac{3}{8\pi}\,\Big[\partial_{t}\,H+2H^2\Big]\,\frac{dG^{-1}(q)}{d q}
\right) =0\,,
\label{eq:Maxwell2}
\end{equation}
for Ricci scalar $R=-6\,(\partial_{t}\,H+2H^2)$ from
our curvature conventions~\cite{KlinkhamerVolovik2008b}.
Solving \eqref{eq:Maxwell2} gives  the integration constant $\mu$,
\begin{equation}
\frac{d\epsilon(q)}{d q}
-\frac{3}{8\pi} \,\Big[\partial_{t}\,H+2H^2\Big] \, \frac{dG^{-1}(q)}{d q}
=\mu \,.
\label{eq:Maxwell3}
\end{equation}
Based on the thermodynamic discussion
of Ref.~\citen{KlinkhamerVolovik2008a},
the integration constant $\mu$ may be called the
``chemical potential,''  where the chemical potential $\mu$ is conjugate
to the conserved quantity $q$ in flat spacetime.
See Ref.~\citen{KlinkhamerVolovik2016b} for further discussion
on the different roles of fundamental and ``conserved'' scalars
for the cosmological constant problem.

\section{Energy exchange between matter and vacuum}
\label{sec:Energy-exchange}

We are, now, interested in the quantum-dissipative energy
exchange between vacuum and matter.
In this case, the chemical potential $\mu$ is no longer constant
and can relax in the evolving universe.  We replace Eq.~(\ref{eq:Maxwell2}) by
\begin{equation}
\partial_{t}\left(\frac{d\epsilon(q)}{d q}
-\frac{3}{8\pi} \,\Big[\partial_{t}\,H+2H^2\Big] \, \frac{dG^{-1}(q)}{d q}\right)
=S \,,
\label{eq:Maxwell4}
\end{equation}
where $S$ is a source term. In a companion paper,\cite{KSV2016}  
we use the following \textit{Ansatz}:                  
\begin{equation}
 S = \Gamma_{q}\,(\partial_{t}\,q)^2 +  \Gamma_{H}\,(\partial_{t}\,H)^2  \,,
\label{S}
\end{equation}
with nonnegative decay constants $\Gamma_{q}$ and $\Gamma_{H}$.
This particular \textit{Ansatz} for $S$ does not discriminate between 
de-Sitter and Minkowski vacua, and the crucial question is whether or not
the Minkowski vacuum is dynamically preferred. 
In Ref.~\citen{KSV2016}, the answer is found to be negative, 
and here we shall use another \textit{Ansatz} for $S$ 
which does prefer the Minkowski vacuum; see Sec.~\ref{Sec:Particle-production}.

The generalized Einstein equation is still valid
and is obtained by variation
of the action \eqref{eq:actionF}  over the metric $g_{\mu\nu}$,
\begin{eqnarray}
&&
\frac{1}{8\pi G(q)}
\left( R_{\mu\nu}-\frac{1}{2}\,R\,g_{\mu\nu}\right)
+\frac{1}{16\pi}\, q\,\frac{d G^{-1}(q)}{d q}\, {R}\,g_{\mu\nu}
\nonumber\\[1mm]
&&
+ \frac{1}{8\pi} \Big( \nabla_\mu\nabla_\nu\, G^{-1}(q) - g_{\mu\nu}\,
\Box\, G^{-1}(q)\Big)
\nonumber\\[1mm]
&&
-\left( \epsilon(q) -q\,\frac{d\epsilon(q)}{d q}\right)g_{\mu\nu}
 +T^\text{\,SM}_{\mu\nu}(\psi) =0\,, \label{eq:EinsteinEquationF}
\end{eqnarray}
where $\Box$ is the invariant d'Alembertian and
$T^\text{\,SM}_{\mu\nu}$ is the  energy-momentum tensor of the
standard-model matter fields $\psi$
(without dependence on $q$ as discussed in Sec.~\ref{sec:Realization}).

\section{Constant--$\boldsymbol{G}$ case}
\label{sec:Constant-G}

For the present article, it suffices to consider the simplest possible
\textit{Ansatz} for the function $G(q)$, namely a constant function,
\begin{equation}
G(q)= G_{N}  \,,
\label{G-const}
\end{equation}
with $G_{N}$ Newton's gravitational constant.
The generalized Einstein and Maxwell equations
from Secs.~\ref{sec:Realization} and \ref{sec:Energy-exchange} become
\begin{eqnarray}\label{eq:Einstein-Maxwell}
\hspace*{-10mm}
&&
\frac{1}{8\pi G_{N}}
\left( R_{\mu\nu}-\frac{1}{2}\,R\,g_{\mu\nu}\right)
=
\rho_{V}(q)\,g_{\mu\nu}
-T^\text{\,SM}_{\mu\nu}\,,
\label{eq:EinsteinEquation2}\\[2mm]
\hspace*{-10mm}
&&
q\,\partial_{t} \left(\frac{d\epsilon(q)}{d q} \right)
=
-\partial_{t}  \rho_{V}(q)
=q\, S\,,
\label{eq:MaxwellEquation2}
\end{eqnarray}
where the source term of the gravitational equation \eqref{eq:EinsteinEquation2}
contains the following vacuum energy density:
\begin{equation}
 \rho_{V}(q)  \equiv   \epsilon(q) -q\,\frac{d\epsilon(q)}{d q}   \,.
\label{Lambdaq}
\end{equation}
Note that the definition \eqref{Lambdaq} also explains
the first equality in \eqref{eq:MaxwellEquation2}.

Consider the spatially-flat FRW universe
with a single homogeneous perfect-fluid matter component (M) and a
homogeneous $q$-field component (V) with pressure $P_{V}=-\rho_{V}$.
The Einstein equation then
gives the following Friedmann equations:%
\bsubeqs\label{eq:Friedmann-eqs} 
\begin{equation}
  3\,H^2 =
 8\pi G_{N}\, (\rho_{V}+\rho_{M})\,,
 \label{eq:Friedmann-eqs-H2}
\end{equation}
\begin{equation}
2\,\partial_{t}\,H=    -8\pi G_{N}\, (\rho_{M}+P_{M}) \,,
\label{eq:Friedmann-eqs-dotH}
\end{equation}
\esubeqs
where the vacuum contribution to the right-hand side of
\eqref{eq:Friedmann-eqs-dotH} cancels out,
because \mbox{$\rho_{V}+P_{V}=0$.}
The evolution equations for the vacuum and matter energy densities are
\bsubeqs\label{eq:Evolution-eqs-V-M}
\beqa
\partial_{t}\,\rho_{V} &=& -q\, S \,,
\label{eq:Evolution-eqs-V}
\\[2mm]
\partial_{t}\,\rho_{M} + 3H\, \Big(P_{M}+\rho_{M}\Big) &=&+ q\, S\,. 
\label{eq:Evolution-eqs-M}
\eeqa
\esubeqs
Now, assume the matter to have a constant equation-of-state parameter,
\begin{equation}
w_{M} \equiv  \rho_{M}/P_{M}=\text{const.}
\label{eq:wM}
\end{equation}
Then,
the following two ordinary differential equations (ODEs)
suffice to determine $q(t)$ and $H(t)$
in a spatially-flat FRW universe:
\bsubeqs\label{eq:two-basic-ODEs}
\begin{equation}
\partial_{t}\left(\frac{d\epsilon(q)}{d q}\right)
=S \,,
\label{eq:Evolution-eqs-V-MaxwellDiss2}
\end{equation}
\begin{equation}
\frac{2}{1+w_{M}}\,\partial_{t}\,H + 3H^2=  8\pi G_{N}\,  \rho_{V}(q)\,,
 \label{eq:Evolution-eqs-V-M-H2dot}
\end{equation}
\esubeqs
with  $\rho_{V}(q)$ from Eq.~(\ref{Lambdaq}) and $S$ from Eq.~(\ref{S})
or otherwise.

The companion paper~\cite{KSV2016} focuses on   
the source term (\ref{S}), which keeps de-Sitter spacetime
stable, whereas the present paper considers another type of
source term which distinguishes Minkowski spacetime. Most importantly,
we do not wish to take some \textit{ad hoc} source term but
will get a term with a clear physical origin.

\section{Particle production and backreaction}
\label{Sec:Particle-production}

Consider the production of massless particles (e.g., gravitons)
by the curved spacetime of an expanding spatially-flat FRW universe
with appropriate boundary conditions on the
matter fields and background~\cite{ZeldovichStarobinsky1977,DobadoMaroto1999,%
BirrellDavies1982,MukhanovWinitzki2007}.
Then, the increase of particle number density is given
by~\cite{ZeldovichStarobinsky1977,DobadoMaroto1999}
\bsubeqs
\beq
\label{eq:nM-dot-eq}
\partial_{t}\,n_{M} \sim R^2 \sim 36\,\Big(\partial_{t}\,H+2H^2\Big)^2 \,,
\eeq
for Ricci scalar $R=-6\,(\partial_{t}\,H+2H^2)$.
The typical particle energy ($E=\hbar\, \omega \geq 0$) is determined
by the Hubble expansion rate
(cf. the discussion on p.~62 of Ref.~\citen{BirrellDavies1982}),
\beq
\label{eq:E-typical}
E_{M} \sim \hbar\,|H| \,,
\eeq
\esubeqs
with $\hbar$ temporarily reinstated.
From \eqref{eq:nM-dot-eq} and \eqref{eq:E-typical},
the standard adiabatic change of the particle energy density then gets
modified by a source term on the right-hand side of the
evolution equation,
\beq
\label{eq:rhoM-dot-eq-rough}
\partial_{t}\,\rho_{M} +3\,H \,(1+w_{M})\,\rho_{M}  \sim \hbar\,|H|\,R^2\,,
\eeq
where $w_M$ equals $1/3$ for the massless particles considered
and, from now on, $\hbar$ will again be set to $1$.
Introducing a dimensionless constant $\gamma$, the matter
evolution equation reads%
\beq
\label{eq:rhoM-dot-eq}
\partial_{t}\,\rho_{M} +4\,H \,\rho_{M}
=(\gamma/36)\, |H|\,R^2 \equiv S_{M}\,,
\eeq
with a further factor $1/36$ inserted for later convenience.

A rough estimate of the coefficient $\gamma$
in  \eqref{eq:rhoM-dot-eq} is based
on the suppressed coefficient in \eqref{eq:nM-dot-eq} calculated
for gravitons~\cite{ZeldovichStarobinsky1977,DobadoMaroto1999}
and the assumed coefficient of unity in \eqref{eq:E-typical}, giving
\beq\label{eq:gammatilde-estimate}
\gamma\;\Big|^\text{\,gravitons} \sim \frac{1}{8\,\pi} \,.
\eeq
An explicit calculation of the $\rho_{M}$ increase by graviton production 
gives precisely the
structure \eqref{eq:rhoM-dot-eq} with correction terms on the right-hand side 
and a calculated coefficient equal to
\beq\label{eq:gammatilde-calculation}
\gamma\;\Big|^\text{\,gravitons} =  \frac{1}{32\,\pi^2}\,,
\eeq
which is a factor $4\,\pi$ smaller
than the naive estimate \eqref{eq:gammatilde-estimate}.
Details of this calculations are relegated to
\ref{app:Energy-density-produced-particles}.

From the Einstein equation \eqref{eq:EinsteinEquation2}
and the contracted Bianchi identities~\cite{Wald1984}
follows the covariant conservation of the total energy-momentum
tensor of the two components considered,
the matter component from the standard-model fields
and the vacuum component from the $q$ field.
In the context of a spatially-flat FRW universe with a single
perfect-fluid matter component (M) and a vacuum component (V)
from the $q$ field, this energy-momentum conservation implies
that the evolution equation of the vacuum energy density
must have precisely
the opposite source term compared to \eqref{eq:rhoM-dot-eq},
\beqa
\label{eq:rhoV-dot-eq}
\partial_{t}\,\rho_{V}  &=&-S_{M}=
-(\gamma/36)\, |H|\,R^2
=
-\gamma\, |H|\,(\partial_{t}\,H+2H^2)^2 \,,
\eeqa
with $\gamma \approx 3.1663\times 10^{-3}$
for graviton production according to \eqref{eq:gammatilde-calculation}.
Hence, the right-hand side of \eqref{eq:rhoV-dot-eq} can be interpreted as
describing the \emph{backreaction} of the particle production
given by Eq.~\eqref{eq:rhoM-dot-eq-rough}; see below for
further discussion.
Note that \eqref{eq:rhoM-dot-eq} and \eqref{eq:rhoV-dot-eq}
are noninvariant under time-reversal,
which is appropriate for dissipative processes~\cite{KSV2016}.

Let us end this section with three general remarks.
First, result \eqref{eq:nM-dot-eq}
relies on being able to define an adiabatic vacuum,
which is possible if the expansion rate
vanishes asymptotically in the past and in the
future~\cite{ZeldovichStarobinsky1977,DobadoMaroto1999,%
BirrellDavies1982}.
For a free massless scalar, the imaginary part of the
effective action, calculated to quadratic order in the curvature,
is given by the spacetime integral of Eq.~(29)
in Ref.~\citen{DobadoMaroto1999}.
For a spatially-flat ($k=0$) FRW universe,
this  spacetime integral reduces to an integral over
the sum of the $R^2$ term and the Gauss--Bonnet term, as
the term with the square of the Weyl tensor vanishes identically.
For an asymptotically-flat $k=0$ FRW
universe, the Gauss--Bonnet term integrates to zero, leaving
the single $R^2$ term.
For interacting quantum fields,
the complete de-Sitter spacetime may give rise to explosive particle
production~\cite{Polyakov2008,Polyakov2010,KrotovPolyakov2011,Polyakov2012}
(used for $q$-theory in Ref.~\citen{Klinkhamer2012})
but the expanding $k=0$  FRW universe considered here does not.
Still, compared to \eqref{eq:rhoM-dot-eq} for free massless
fields in an expanding $k=0$  FRW universe, there may be
a somewhat enhanced particle production due to particle self-interactions~\cite{KrotovPolyakov2011}.

Second, it is well-known that the gravitational backreaction
of quantum matter fields is a subtle problem, which is not completely
solved~\cite{BirrellDavies1982,MukhanovWinitzki2007,Wald1984}. Our description
is the simplest possible: keep unchanged the form of the
energy-momentum tensor on the right-hand sides of
Eqs.~\eqref{eq:Friedmann-eqs-H2} and \eqref{eq:Friedmann-eqs-dotH}
and modify both the evolution equation
of the matter component \eqref{eq:rhoM-dot-eq}
and the evolution equation of the vacuum component \eqref{eq:rhoV-dot-eq}.
Other changes are certainly to be expected
(e.g., vacuum polarization effects), but
our minimal description suffices for an exploratory study.

Third, a heuristic explanation of the modified evolution equations
is as follows.
Start with a spatially-flat FRW universe having a nonvanishing energy
density $\rho_{M}\ne 0$ from relativistic particles  ($w_{M} =1/3$)
and a vanishing vacuum energy density $\rho_{V}(\overline{q}) = 0$
from an appropriate $q$-field value $\overline{q}$.
Then, $q$ stays at the value $\overline{q}$  and $\rho_{M}$
dilutes by expansion [$\rho_{M}\propto 1/a(t)^4$]
without extra particle production.
Intuitively, it is clear that a Hubble expansion driven solely by
massless particles, $H \equiv (\partial_{t}\,a)/a = \half\,(t-t_0)^{-1}$,
does not create more massless particles.
If, now, the Hubble expansion is modified [$H \ne \half\,(t-t_0)^{-1}$]
by having $\rho_{V}(q) > 0$,
then there will be particle production as
$R \propto (\partial_{t}\,H+2\, H^2) \ne 0$.
The ``agent'' of this particle production is the
nontrivial vacuum field $q-\overline{q}$. Ultimately,
the energy produced in particles must come from the
agent (here, $q-\overline{q}$) responsible for the
modification of the Hubble expansion, $H \ne \half\,(t-t_0)^{-1}$.
This discussion is analogous to that of the Unruh effect:
the energy for the thermal radiation heating up
the uniformly-accelerated detector ultimately comes
from the agent responsible for the acceleration of the detector
(cf. p. 55 of Ref.~\citen{BirrellDavies1982} and p. 108 of Ref.~\citen{MukhanovWinitzki2007}). Another analogy
is with Schwinger pair creation in a uniform static electric   field~\cite{KrotovPolyakov2011,DobadoMaroto1999,MukhanovWinitzki2007}:
the energy for the created particles ultimately comes for the
agent responsible for the electric field.

\section{$\boldsymbol{Q}$-theory model of vacuum-energy decay}
\label{Sec:Q-theory-model-of-vacuum-energy-decay}

Henceforth, we use the same dimensionless variables
as in Ref.~\citen{KlinkhamerVolovik2008b}, effectively
obtained by rescaling with appropriate powers of the
Planck energy $E_{P} \equiv
\sqrt{\hbar\,  c^5/G_{N}} \approx 1.22 \times 10^{19}\,\text{GeV}$
and denoted by lower-case letters. Specifically, we have
the dimensionless time $\tau$ and the dimensionless Hubble
parameter $h$ (taken to be positive).
The 4-form field strength \eqref{eq:Fdefinition2}
gives rise to the pseudoscalar field $q$ of mass dimension 2
and rescaling this field $q$
(denoted $F$ in Ref.~\citen{KlinkhamerVolovik2008b})
produces the dimensionless variable $f$.
The overdot in the differential equations below
will denote differentiation with respect to
$\tau$ and the prime differentiation with respect to
$f$.

We now present the $q$-theory equivalent of
the source term found in Sec.~\ref{Sec:Particle-production},
which physically corresponds to the backreaction from
particle production by the curved spacetime.
In addition, we allow for a  dimensionless cosmological constant
$\lambda\equiv \Lambda/(E_{P})^4$ in the dimensionless
energy density $\epsilon(f)$.
This additional cosmological constant $\Lambda$ represents
the effects from zero-point energies of the matter quantum
fields~\cite{Weinberg1988}
and cosmic phase transitions~\cite{KirzhnitsLinde1976}.

Specifically, the generalized Maxwell
equation \eqref{eq:Evolution-eqs-V-MaxwellDiss2}
with the specific source term  from \eqref{eq:rhoV-dot-eq}
and the generalized Friedmann
equation \eqref{eq:Evolution-eqs-V-M-H2dot} for $w_M=1/3$ give
\bsubeqs\label{eq:2eqsFRWdim4}
\beqa
f\;\dot{f} \;\epsilon^{\prime\prime}&=&
\gamma\,|h|\,\Big(\dot{h}+2\, h^2\Big)^2 \,,
\label{eq:MaxwellFRWdim4}
\\[2mm]
\dot{h}+ 2\, h^2&=& 2\,r_{V}\,,
\label{eq:EinsteinFRWdim4}
\eeqa
\esubeqs
with the dimensionless gravitating vacuum energy density
\bsubeqs\label{eq:CCP-rV-definition-epsilon-Ansatz}
\beqa
\label{eq:CCP-rV-definition}
r_{V}(f)  &=&  \epsilon(f) - f\,\epsilon^{\prime}(f)
\eeqa
and the particular \textit{Ansatz} function
\beqa
\label{eq:CCP-epsilon-Ansatz}
\epsilon(f) &=&  \lambda+ f^2 + 1/f^2\,.
\eeqa
\esubeqs
The  \textit{Ansatz} \eqref{eq:CCP-epsilon-Ansatz}
has two important properties:
first, the corresponding values of $r_{V}=\lambda-f^2+3/f^2$
range over $(-\infty,\infty)$
for $f^2 \in (0,\infty)$ and any finite value of $\lambda$;
second, the corresponding vacuum compressibility~\cite{KlinkhamerVolovik2008a}
$\chi\equiv \big(f^2\, d^2\epsilon/df^2\big)^{-1}$ is positive for
any $f^2 \in (0,\infty)$. 
Other \textit{Ans\"{a}tze} for $\epsilon(f)$ are certainly possible.

Note that, strictly speaking,
the ODEs \eqref{eq:2eqsFRWdim4} can be written
solely in terms of $r_{V}(\tau)$ and $h(\tau)$, without need of the
$q$-type field $f(\tau)$. But this is only because of the
special case considered, $G(q)=\text{const}$.
For generic $G(q)$, the $q$ field appears explicitly on the left-hand
side of the generalized Maxwell equation \eqref{eq:Maxwell4}.
Moreover, precisely $q$ theory in the four-form realization gives
rise to the $\rho_{V}$ evolution equation as a field equation,
namely, the generalized Maxwell equation \eqref{eq:Maxwell2}
for the classical theory.
For these reasons, it is appropriate to speak of
a $q$-theory model of vacuum-energy decay.

The two ODEs \eqref{eq:MaxwellFRWdim4} and \eqref{eq:EinsteinFRWdim4}
are to be solved simultaneously
and the matter energy density is obtained from the solutions
$\overline{f}(\tau)$ and $\overline{h}(\tau)$ by
\beqa
r_{M}&=&\overline{h}^2 - r_{V}\big(\,\overline{f}\,\big)\,.
\label{eq:rM-from-h-f}
\eeqa
The numerical solutions of the ODEs \eqref{eq:2eqsFRWdim4} will
be presented in the next section and some analytic results will be
given in \ref{app:Analytic-results-ODEs}.

\section{Minkowski-vacuum attractor}
\label{Sec:Minkowski-vacuum-attractor}

\begin{figure*}[t]  
%
%
\vspace*{-0cm}
\begin{center}
\includegraphics[width=.75\textwidth]{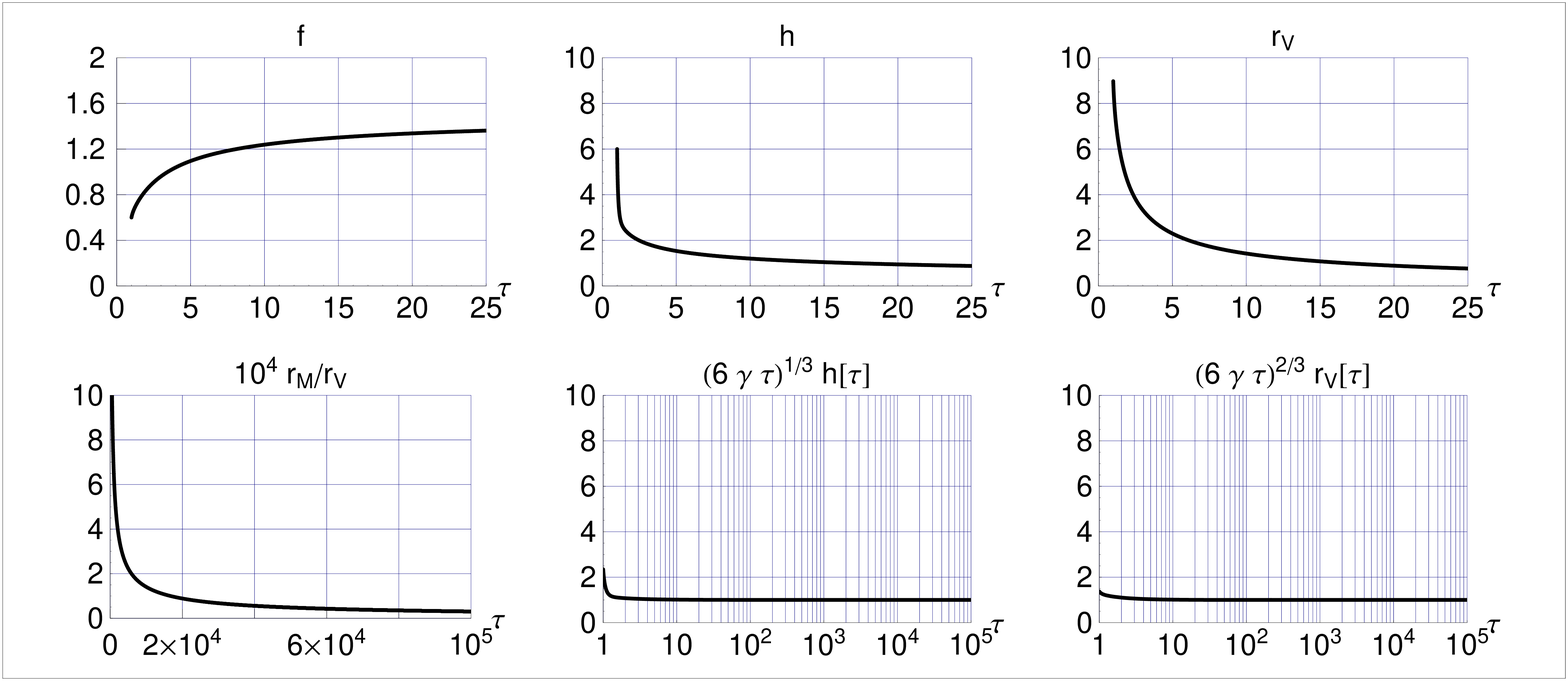}
\end{center}
\vspace*{-1mm}
\caption{Numerical solution of the  ODEs \eqref{eq:2eqsFRWdim4}
with auxiliary
functions \eqref{eq:CCP-rV-definition-epsilon-Ansatz}.
The model parameters are
$\{\lambda,\gamma\}=\{1,\,1/100\}$ and the boundary conditions
at $\tau=1$ are $\{h(1),f(1)\}=\{6,\, 3/5\}$.
The initial energy densities  are
$\{r_{V}(1),r_{M}(1)\}=\{ 8.97333,\, 27.0267\}$.
The value of the vacuum energy density at $\tau=10^5$ is
$r_{V}(10^{5})= 3 \times 10^{-3}$.
}
\label{fig:FlatFRWuniverse-relmat-short-constantG-CCP-lambda-pos}
\vspace*{1mm}
\begin{center}
\includegraphics[width=.75\textwidth]{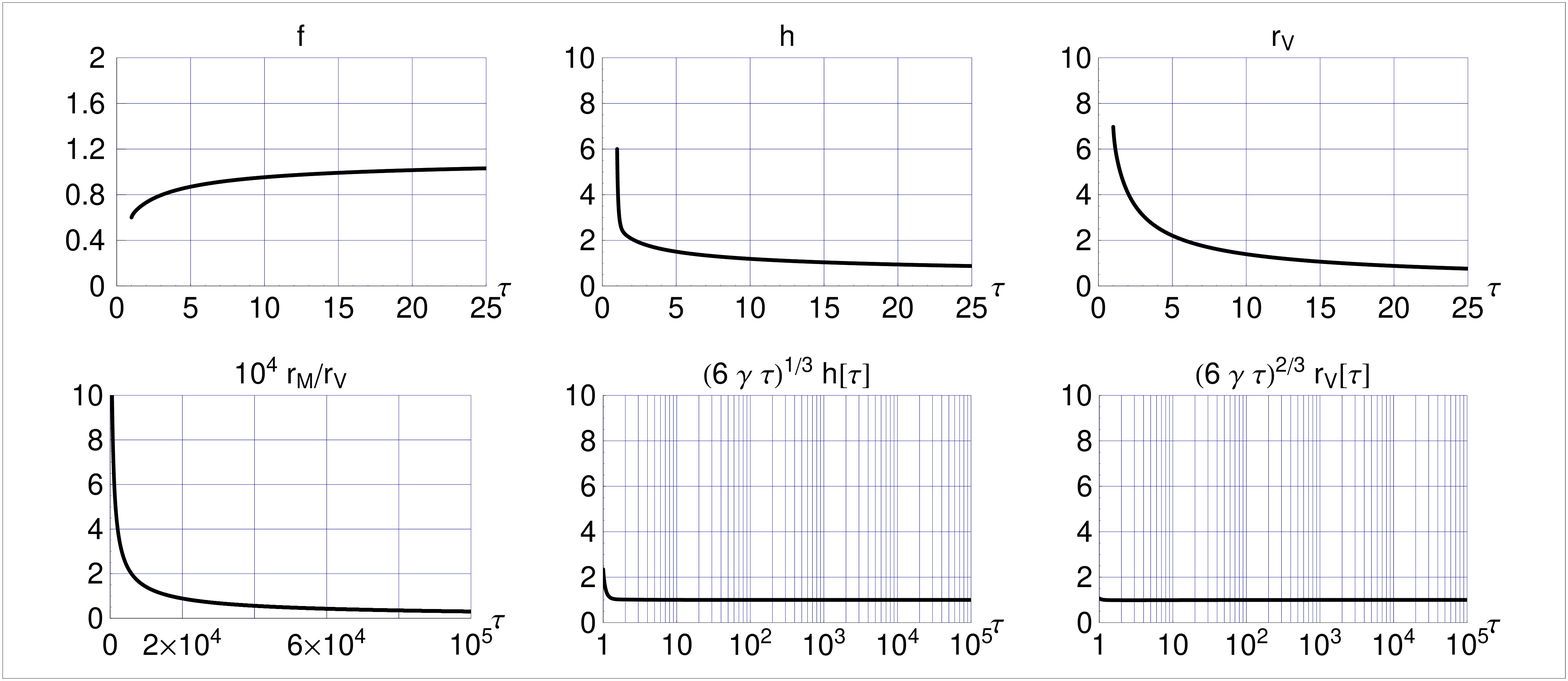}
\end{center}
\vspace*{-1mm}
\caption{Same parameters and boundary conditions
as in Fig.~\ref{fig:FlatFRWuniverse-relmat-short-constantG-CCP-lambda-pos},
except for a different value of the  cosmological constant,
$\lambda=-1$. The initial energy densities are
$\{r_{V}(1),r_{M}(1)\}=\{ 6.97333,\, 29.0267\}$.
The value of the vacuum energy density at $\tau=10^5$ is
$r_{V}(10^{5})= 3 \times 10^{-3}$.
}
\label{fig:FlatFRWuniverse-relmat-short-constantG-CCP-lambda-neg}
\vspace*{1mm}
\begin{center}
\includegraphics[width=.75\textwidth]{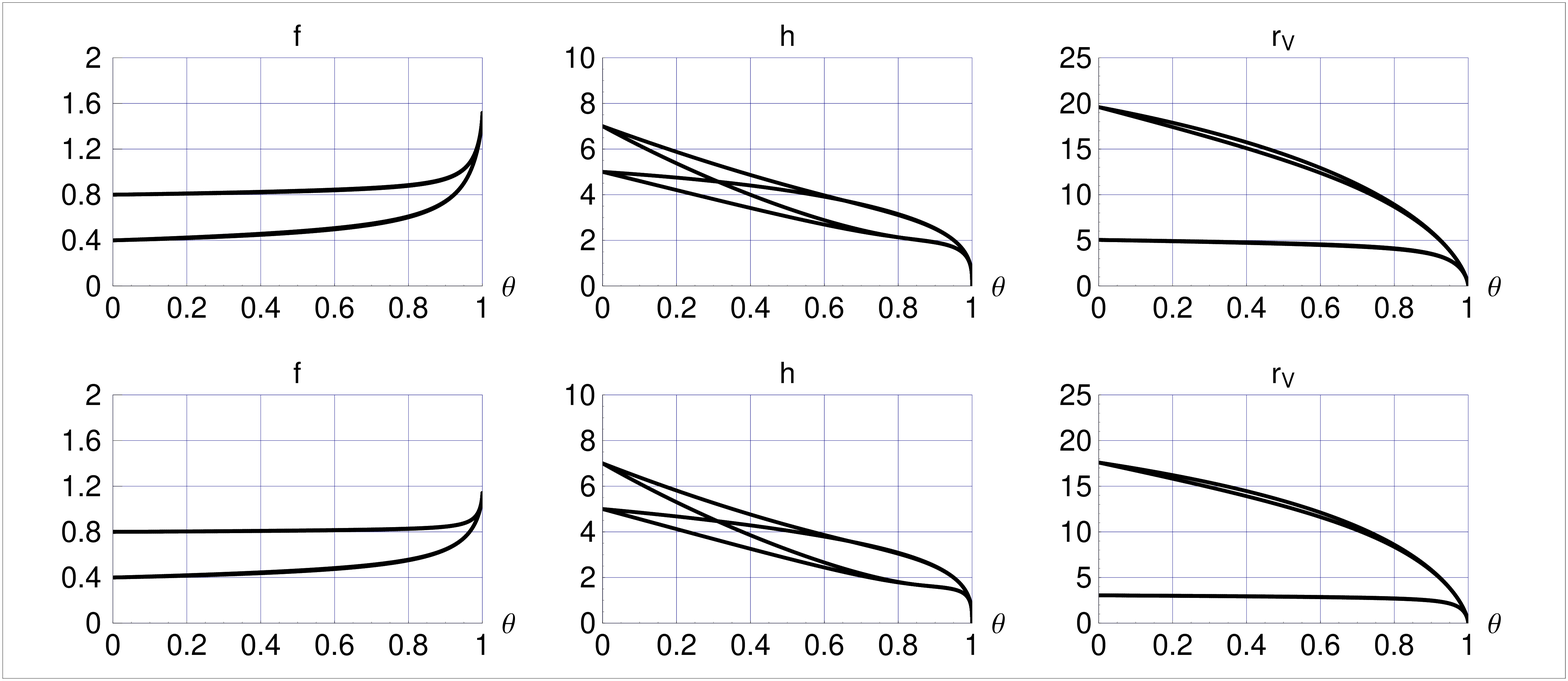}
\end{center}
\vspace*{-1mm}
\caption{Numerical solutions of the ODEs \eqref{eq:2eqsFRWdim4}
with auxiliary functions \eqref{eq:CCP-rV-definition-epsilon-Ansatz}
for $\lambda=1$ (top row) and $\lambda=-1$ (bottom row).
The numerical value of the decay parameter $\gamma$ is $1/100$.
The functions $f,h$ and $r_{V}[f]$
are plotted versus the compactified time coordinate
$\theta \equiv (\tau-1)/(\tau+\tau_\text{mid}-2)$ with $\tau_\text{mid}=11/10$.
Four sets of boundary conditions at $\theta=0$ are used:
$\{h(0),f(0)\}=\{6\pm 1,\, 3/5 \pm 1/5\}$.
}
\label{fig:FlatFRWuniverse-relmat-short-constantG-CCP-attractor}
\end{figure*}

From the ODEs  \eqref{eq:MaxwellFRWdim4} and \eqref{eq:EinsteinFRWdim4}
follows that the curves for  $r_{V}(\tau)$ and  $h(\tau)$
are monotonically decreasing, provided that
the decay constant $\gamma$ is positive and that
$r_{M}(\tau)= h(\tau)^2 - r_{V}[f(\tau)]$ stays positive for all
values of $\tau$. Numerical solutions are given in
Figs.~\ref{fig:FlatFRWuniverse-relmat-short-constantG-CCP-lambda-pos}
and \ref{fig:FlatFRWuniverse-relmat-short-constantG-CCP-lambda-neg}
for positive and negative cosmological constants, $\lambda=\pm 1$.
Similar numerical results have been obtained for $\lambda=0$.
These results show that the $r_{V}=0$ Minkowski vacuum
is approached without need of fine-tuning: the required asymptotic
values of the $q$-type field $f$ are generated dynamically
(see the top-left panels of
Figs.~\ref{fig:FlatFRWuniverse-relmat-short-constantG-CCP-lambda-pos}
and \ref{fig:FlatFRWuniverse-relmat-short-constantG-CCP-lambda-neg}).

Note that the decay constant used for these numerical
results, $\gamma=10^{-2}$, has a realistic order of magnitude
(see Sec.~\ref{Sec:Particle-production}).
Similar numerical results have been obtained for other
values of the decay coupling constant, ranging from $\gamma=1$
down to $\gamma=10^{-3}$.

The numerical results establish the existence of the
$r_{V}=0$ \emph{attractor}
for $\lambda\in \{-1,\,0,\,+1\}$ with a finite domain of
boundary conditions
$\{h(1),f(1)\}$ $=$ $\{6\pm 1,\, 0.6\pm 0.2\}$;
see Fig.~\ref{fig:FlatFRWuniverse-relmat-short-constantG-CCP-attractor}.
The actual domain of attraction for $|\lambda| \leq 1$
may be larger than this rectangle
(see also the last paragraph of \ref{app:Analytic-results-ODEs}
for further discussion of the attractor domain).

The heuristic understanding for the appearance of an
attractor is as follows.
The left-hand side of \eqref{eq:MaxwellFRWdim4} equals $-\dot{r}_{V}$
according to \eqref{eq:CCP-rV-definition}.
Using \eqref{eq:EinsteinFRWdim4} for $h^2$,
the ODE \eqref{eq:MaxwellFRWdim4} then
reads $\dot{r}_{V}= - 4\gamma\, |h|\, (r_{V})^2 
+ \text{(rest-terms)}$.  
This is not quite the structure relevant for
the Poincar\'{e}--Lyapunov theorem
(given as Theorem 66.2 in Ref.~\citen{Hahn1968}
and Theorem 7.1 in Ref.~\citen{Verhulst1996}), but does indicate 
a weak approach to the $r_{V}=0$ asymptote.
In fact, the asymptotic behavior
from  \eqref{eq:2eqsFRWdim4} is given by
\bsubeqs\label{eq:asymptotic-h-f}
\beqa
h(\tau) &\sim& (6\,\gamma\,\tau)^{-1/3} \,,
\label{eq:asymptotic-h}
\\[2mm]
r_{V}(\tau) &\sim& (6\,\gamma\,\tau)^{-2/3}\,.
\label{eq:asymptotic-f}
\eeqa
\esubeqs

Note that the attractor  behavior found here
is qualitatively similar to the one of Dolgov theory~\cite{Dolgov1985,Dolgov1997}
shown numerically in Fig.~2 of Ref.~\citen{KlinkhamerVolovik2010}
and proven mathematically in App. A of Ref.~\citen{EmelyanovKlinkhamer2011}.
But the Dolgov theory as such
ruins the standard Newtonian dynamics~\cite{RubakovTinyakov1999}
and needs to be modified
significantly~\cite{EmelyanovKlinkhamer2011-CCP1-NEWTON,EmelyanovKlinkhamer2011}.

Three final remarks are in order.
First, the asymptotic decay of the Hubble parameter
is slow and gives rise to an
\emph{inflationary behavior}~\cite{Starobinsky1980,Guth1981,Linde1983}
of the particle horizon,
\beqa
d_\text{hor}(\tau)
&\equiv&
d_{1} + a(\tau)\,\int_{1}^{\tau}\, \frac{d\tau'}{a(\tau')}
\sim
\exp\left[\frac{3}{2}\;\frac{\left(  \tau^{2/3} -1 \right) }
  {(6\,{\gamma})^{1/3}}\right]\,,
\eeqa
where $a(\tau)$ is the scale factor defined by
$h= \dot{a}/a$ for $h(\tau)$ given by \eqref{eq:asymptotic-h}
and $d_{1}$ is the contribution from times before $\tau=1$
(a radiation-dominated universe with an initial singularity at
$\tau=0$ gives $d_{1}=2$).
Note that this inflationary behavior
holds only  as long as the particle production is given by the
$|H|\,R^2$ term in \eqref{eq:rhoM-dot-eq-rough}, whose
dominance over other contributions needs to be verified
[see the discussion in \ref{app:Energy-density-produced-particles}].

Second,  the same type of slow asymptotic decay,
$h(\tau) \propto \tau^{-1/3}$ and $r_{V}(\tau) \propto \tau^{-2/3}$,
has also been found in a nonconstant--$G$ model with
\textit{Ans\"{a}tze} for $G(f)$ and $\epsilon(f)$
from Ref.~\citen{KlinkhamerVolovik2008b}, but now with an arbitrary
cosmological constant $\lambda$ added to $\epsilon(f)$.

Third, returning to the constant--$G$ model considered here,
it needs to be emphasized that we remain within the framework
of standard general relativity
(with certain quantum effects of the matter fields included,
as discussed in Sec.~\ref{Sec:Particle-production}).
Moreover, there are essentially no free parameters in the
equation system \eqref{eq:2eqsFRWdim4},
as the decay constant $\gamma$ has been
calculated to be of order $3\times 10^{-3}$ for gravitons,
according to \eqref{eq:gammatilde-calculation}
and \eqref{eq:intermediate-rho-gravitons-main-result-local+rest}.

\section{Conclusion}
\label{Sec:Conclusion}

In this article, we have again addressed the
cosmological constant problem, which can be formulated as follows:
how can it be that the vacuum state does not have an effective
cosmological constant $\Lambda$
(or gravitating vacuum energy density $\rho_{V}=\Lambda$
and pressure $P_{V}=-\Lambda$)
with an energy scale of the order of the known energy scales
of elementary particle physics? A particular
adjustment-type solution of the cosmological constant problem
involves so-called $q$ fields,
which are (pseudo-)scalar composites of
higher-spin fields (for example, $q$ as a pseudoscalar composite from
the field strength  $F_{\kappa\lambda\mu\nu}$
and the metric $g_{\mu\nu}$).

The $q$-theory framework serves as the proper tool for
studying physical processes
related to the quantum vacuum. It describes, in particular, the relaxation
of the vacuum energy density (effective cosmological constant) as the backreaction
of the deep vacuum to different types of perturbations, such as the Big Bang, inflation, cosmological phase transitions, and vacuum instability
in gravitational or other backgrounds.

By considering the energy exchange between this $q$ field
in the four-form-field-strength realization
and massless particles produced by the spacetime
curvature, we have found that 
a Planck-scale cosmological constant $\Lambda$ of arbitrary sign
is cancelled by the $q$-field dynamics without fine-tuning. 
As mentioned previously, this cancellation occurs
within the realm of standard general relativity.

The Minkowski vacuum with $\rho_{V}=-P_{V}=0$ appears as
an attractor of the dynamical equations
(see Fig.~\ref{fig:FlatFRWuniverse-relmat-short-constantG-CCP-attractor}
and \ref{app:Analytic-results-ODEs}). As the approach
to the Minkowski vacuum is rather slow, there occurs an
inflationary behavior of the particle horizon, provided
the nature of the particle production does not change significantly.
The existence of such an inflationary phase
(possibly before the start of the ``standard'' matter-dominated
FRW universe) also requires
that there are no other, more efficient dissipation processes
than particle production by spacetime curvature.

\begin{appendix}
\section{Energy density of produced particles}
\label{app:Energy-density-produced-particles}

In this appendix, we calculate the energy
density of gravitons produced by spacetime curvature,
directly following the number-density calculation of
Zel'dovich and Starobinsky~\cite{ZeldovichStarobinsky1977}.
We refer to their paper~\cite{ZeldovichStarobinsky1977} for further
details and use the same notation.
See also the textbooks~\cite{BirrellDavies1982,MukhanovWinitzki2007}
for a general discussion of particle production.

From the Bogoliubov coefficient $\beta_{k}$, the final energy density of
produced massless gravitons is given by%
\beqa\label{eq:rho-gravitons-start}
\rho_{M,\,\text{gravitons}} &=&
2\; (2\pi)^{-3}\;a^{-4}\;
\int d^{3}k\;|\vec{k}|\; |\beta_{k}|^2 \;
\exp\left[-\epsilon\,|\vec{k}|^2\right] \,.
\eeqa
Compared to the expression for the number density
[given by Eq.~(6) in Ref.~\citen{ZeldovichStarobinsky1977}
for massless real scalars], there are
several different factors in \eqref{eq:rho-gravitons-start}:
the first factor $2$
is for the two helicity states of the graviton, the
factor $a^{-4}$ contains an extra factor $a^{-1}$ for the
redshift of the energy, the integrand of the $k$-integral has the
energy factor $|\vec{k}|$ and, finally, we have added a positive
regulator $\epsilon$ which is taken to $0$ at the end of the
calculation.

The calculation will use the conformal time $\eta$, defined in the
standard way by $\eta=\int^{t} d\widetilde{t}/a(\widetilde{t})$.
The wave equation for gravitons with comoving   
wave number $k$
is given by~\cite{ZeldovichStarobinsky1977}%
\bsubeqs\label{eq:wave-eq-V-def}
\beqa
\label{eq:wave-eq}
\chi_{k}''(\eta) + k^2\,\chi_{k}(\eta) &=& V(\eta)\,\chi_{k}(\eta) \,,
\\[2mm]
\label{eq:V-def}
V(\eta)  &\equiv& a''(\eta)/a(\eta) \,,
\eeqa
\esubeqs
where the prime denotes differentiation with respect to $\eta$.

Now take the $\beta_{k}$ expression
[given by Eq.~(5) in Ref.~\citen{ZeldovichStarobinsky1977}]
evaluated for the potential term $V$ from \eqref{eq:V-def}
and for the zeroth-order wave function $\chi_{k}^{(0)}(\eta)=\exp[-ik\eta]$.
Inserting this expression for $\beta_{k}$ into
\eqref{eq:rho-gravitons-start} gives the
following triple integral:
\beqa\label{eq:rho-gravitons-triple-integral}
\rho_{M} &=&
2\; (2\pi)^{-3}\;a^{-4}\;
\int_{-\infty}^{\infty} d\eta_{1}\;\int_{-\infty}^{\infty} d\eta_{2}\;
\int_{0}^{\infty} d k\;
\nonumber\\[1mm]&&
\times
\pi\;k\; \exp[2ik(\eta_{1}-\eta_{2})] \;\exp[-\epsilon\,k^2]\;
V(\eta_{1})\;V(\eta_{2}) \,.
\eeqa
The calculation of this multiple integral is somewhat subtle,
as there is a quadratic divergence
of the $k$ integral if the exponentials are omitted.


Introducing new coordinates
\beq\label{eq:eta-pm-def}
\eta_{\pm} \equiv \frac{\eta_{1} \pm \eta_{2}}{2}\,,
\eeq
the evaluation of \eqref{eq:rho-gravitons-triple-integral}
gives the main result of this appendix,
\beqa\label{eq:rho-gravitons-main-result}
\rho_{M,\,\text{gravitons}}
&=&
\frac{1}{32\,\pi^2}\; a^{-4}\;
\int_{-\infty}^{\infty} d\eta_{+}\;\int_{-\infty}^{\infty} d\eta_{-}\;
\nonumber\\[1mm]&&
\times
\frac{1}{(\eta_{-})^2}\;
\Big[V(\eta_{+})\;V(\eta_{+})
-V(\eta_{+}+\eta_{-})\;V(\eta_{+}-\eta_{-})\Big]\,,
\eeqa
with $V$ defined by \eqref{eq:V-def}.
The integral over $\eta_{-}$ in \eqref{eq:rho-gravitons-main-result}
produces a type of damped auto-correlation function of $V(\eta)$.
Equation \eqref{eq:rho-gravitons-main-result} has a well-defined
integrand (also at $\eta_{-}=0$) and improves upon
expression (5.122) of Ref.~\citen{BirrellDavies1982}, specialized to
the isotropic background.

Assuming that the main contribution to the integral
over $\eta_{-}$ in \eqref{eq:rho-gravitons-main-result}
comes from an interval $\Delta\eta>0$
and Taylor expanding  the second term in the square
brackets gives the following rough estimate:
\beq\label{eq:rho-gravitons-main-result-tmp1}
\widehat{\gamma}\; a^{-4}\;
\int_{-\infty}^{\infty} d\eta_{+}\;\Big[\Delta\eta\;V'(\eta_{+})\;V'(\eta_{+})\Big]\,,
\eeq
with $\widehat{\gamma} \equiv 1/(32\,\pi^2)$.
If we now set $\Delta\eta \; V'(\eta_{+}) \sim \Delta V \sim V$,
then the result is
\beq\label{eq:rho-gravitons-main-result-tmp2}
\widehat{\gamma}\; a^{-4}\;
\int_{-\infty}^{\infty} d\eta_{+}\;\Big|V(\eta_{+})\;V'(\eta_{+})\Big|\,.
\eeq
Taking only the second term in $V'=(a'''/a)-(a''/a)\,(a'/a)$,
recalling that $6\,V\equiv 6\,a''/a \subset a^2\, R$,
and changing back to the coordinate time $t$, the resulting
expression for the energy density of produced gravitons at $t=\infty$
is
\beqa\label{eq:final-rho-gravitons-main-result-local+rest}
\hspace*{-5mm}
\rho_{M,\,\text{gravitons}}(\infty) &=&
\frac{1}{1152\,\pi^2}\;
\big[a(\infty)\big]^{-4}
\;\int_{-\infty}^{\infty} d \widetilde{t}\;\;
a^4(\widetilde{t}) \;|H(\widetilde{t})|\;R^2(\widetilde{t}) \;+\;\cdots \,,
\eeqa
with the Hubble parameter $H=a'/a^2=(\partial_{t}\,a)/a$ and
the ellipsis standing for higher-derivative local terms
(single integrals) and further nonlocal terms (double integrals).
Also note that we get the absolute value $|H|$  in \eqref{eq:final-rho-gravitons-main-result-local+rest}
because \eqref{eq:rho-gravitons-main-result-tmp1}
has a manifestly positive integrand.

For an alternative way to arrive at the local term
in \eqref{eq:final-rho-gravitons-main-result-local+rest} return to
the main result \eqref{eq:rho-gravitons-main-result}.
Observe that if the conformal-time correlation length of
$V$ is $\Delta\eta>0$, then the integral over $\eta_{-}$
gives approximately $(\Delta\eta)^{-1}\,V^2$ $=$
$a^4\,(\Delta\eta)^{-1}\,(a''/a^3) \,(a''/a^3)$.
If we now set $a\,\Delta\eta \sim \Delta t \sim |H|^{-1}$,
we have for the integrand of the $\eta_{+}$ integral in
\eqref{eq:rho-gravitons-main-result} approximately $a^5\,|H|\,R^2$,
which gives \eqref{eq:final-rho-gravitons-main-result-local+rest}.

As the integrand of the integral on the right-hand side of \eqref{eq:final-rho-gravitons-main-result-local+rest} is nonnegative,
we can obtain an equation for the energy density at time $t$
by taking $t$ as the upper limit on the integral and as the argument
of the factor $a^{-4}$,
\beqa\label{eq:intermediate-rho-gravitons-main-result-local+rest}
\rho_{M,\,\text{gravitons}}(t) &=&
\frac{1}{1152\,\pi^2}\;  \big[a(t)\big]^{-4}
\; \int_{-\infty}^{t} d \widetilde{t}\;\;a^4(\widetilde{t}) \;
|H(\widetilde{t})|\;R^2(\widetilde{t}) \;+\;\cdots \,.
\eeqa
From \eqref{eq:intermediate-rho-gravitons-main-result-local+rest}
follows that the local change of the energy density is
given by \eqref{eq:rhoM-dot-eq} with additional terms
appearing on the right-hand side.

\section{Analytic results}
\label{app:Analytic-results-ODEs}

The ODEs \eqref{eq:2eqsFRWdim4}
can be solved analytically if the following variable is used:
\beq
\xi \equiv \ln[a(\tau)]\,,
\eeq
where `$\ln$' is the natural logarithm and
$a(\tau)$ the FRW cosmic scale factor.
Since the Hubble parameter $h$ (assumed positive)
corresponds to the rate of change
of the scale factor, $h(\tau)=\dot{a}/a$,
we have that $h^{-1}\,d/d\tau$ equals $d/d\xi$.
Furthermore, we take the time derivative of ODE \eqref{eq:EinsteinFRWdim4}
and write the resulting second-order ODE as a
pair of first-order ODEs. The three first-order ODEs are then
\bsubeqs\label{eq:3eqsFRW}
\beqa
\frac{d}{d\xi}\, r_{V}   &=&  -4\, \gamma\, r_{V}^2\,,
\label{eq:3eqsFRW-rV-prime}
\\[2mm]
\frac{d}{d\xi}\, k   &=& -2\, \gamma\, k^2 \,,
\label{eq:3eqsFRW-k-prime}
\\[2mm]
h \frac{d}{d\xi}\, h +2\,h^2    &=&  k \,,
\label{eq:3eqsFRW-h-prime}
\eeqa
\esubeqs
where $k=\dot{h} +2\,h^2$ is proportional to the Ricci scalar $R$
of the spatially-flat FRW universe considered.

The solution of the ODEs \eqref{eq:3eqsFRW} with positive
Hubble parameter $h$ is given by
\bsubeqs\label{eq:3eqsFRW-sol}
\beqa
r_{V}(\xi)   &=& \frac{1}{4\gamma\,(\xi-\xi_{0})} \,,
\label{eq:3eqsFRW-rV-sol}
\\[2mm]
k(\xi)    &=& \frac{1}{2\gamma\,(\xi-\xi_{0})}\,,
\label{eq:3eqsFRW-k-sol}
\\[2mm]
h(\xi)   &=&
\left(
\frac{C_{1}+\exp[4\,\xi_{0}]\; \text{Ei}\,[4\,\xi-4\,\xi_{0}] }
     {\gamma\,\exp[4\,\xi]}
\right)^{1/2} \,,
\label{eq:3eqsFRW-h-sol}
\eeqa
\esubeqs
with two integration constants, $\xi_{0}$ and $C_{1}$, and  
the exponential integral function $\text{Ei}\,[y]$, for $y>0$
defined by~\cite{AbramowitzStegun}
\beq
 \text{Ei}\,[y] \equiv - \text{P} \int_{-y}^{+\infty} dx\;\frac{\exp[-x]}{x}\,,
\eeq
where `$\text{P}$' stands for the principal value of the integral.
Remark that the solutions of \eqref{eq:3eqsFRW} have three integration constants, but the original ODE \eqref{eq:EinsteinFRWdim4}
fixes the integration constants in
\eqref{eq:3eqsFRW-rV-sol} and \eqref{eq:3eqsFRW-k-sol} to be equal.

In principle, it is possible to obtain the function
$\xi(\tau)$
by replacing $h(\xi)$ on the left-hand side of \eqref{eq:3eqsFRW-h-sol}
by $d\xi/d\tau$ and solving the resulting first-order ODE
for $\xi(\tau)$.
However, this solution $\xi(\tau)$ is difficult to obtain
analytically and numerical methods are more appropriate
(cf. Sec.~\ref{Sec:Minkowski-vacuum-attractor}).

Regarding the size of the attractor domain of initial boundary conditions
$\{h(1),f(1)\}$ as discussed in Sec.~\ref{Sec:Minkowski-vacuum-attractor}, 
the solution \eqref{eq:3eqsFRW-sol} gives the answer: $f(1)$ must be 
such as to make $r_{V}(1)$ nonnegative, where $r_{V}[f]$ is given by
\eqref{eq:CCP-rV-definition} 
for the \textit{Ansatz} \eqref{eq:CCP-epsilon-Ansatz}, 
and $h(1)$ must also be nonnegative.

\end{appendix}

\section*{Acknowledgment}

The work of GEV has been supported by
the Academy of Finland (project no. 284594).



\end{document}